\documentclass[pra,aps,onecolumn,floatfix,
               showpacs,noshowpacs,
               balancelastpage,notitlepage,
               longbibliography,preprintnumbers,
               superscriptaddress, final]{revtex4-1}

\usepackage{amsthm}
\usepackage{amsmath}
\usepackage{amssymb}
\usepackage{mathtools}
\usepackage{physics}
\usepackage[table,xcdraw]{xcolor}
\usepackage{graphicx}
\usepackage{grffile}
\usepackage[T1]{fontenc}
\usepackage{csquotes}
\usepackage{longtable}
\usepackage[pdftex, pdftitle={Article}, pdfauthor={Author}]{hyperref}
\usepackage{epsfig}
\usepackage{multirow}
\usepackage{bm}
\usepackage{color}
\usepackage{url}

\usepackage[table,xcdraw]{xcolor}

\newcommand{\eF}{\ensuremath e_{\mathrm{F}}}
\newcommand{\kF}{\ensuremath k_{\mathrm{F}}}

\renewcommand{\abs}[1]{\lvert{#1}\rvert}            

\newcommand\respadd[1]{#1}

\begin{document}

\title{Generation and decay of Higgs mode in a strongly interacting Fermi gas}
\author{Andrea Barresi\email{andrea.barresi.dokt@pw.edu.pl}}
\affiliation{Faculty of Physics, Warsaw University of Technology, Ulica Koszykowa 75, 00-662 Warsaw}
\author{Antoine Boulet}
\affiliation{Faculty of Physics, Warsaw University of Technology, Ulica Koszykowa 75, 00-662 Warsaw}
\affiliation{ISMANS CESI, 44 avenue Fr\'{e}d\'{e}ric Auguste Bartholdi, 72000 Le Mans, France}
\author{Gabriel Wlaz\l{}owski\email{gabriel.wlazlowski@pw.edu.pl}}
\email[Corresponding author: ]{gabriel.wlazlowski@pw.edu.pl}
\affiliation{Faculty of Physics, Warsaw University of Technology, Ulica Koszykowa 75, 00-662 Warsaw}
\affiliation{Department of Physics, University of Washington, Seattle, Washington 98195-1560, USA}
\author{Piotr Magierski\email{piotr.magierski@pw.edu.pl}}
\affiliation{Faculty of Physics, Warsaw University of Technology, Ulica Koszykowa 75, 00-662 Warsaw}
\affiliation{Department of Physics, University of Washington, Seattle, Washington 98195-1560, USA}
\begin{abstract}
We investigate the life cycle of the large amplitude Higgs mode in strongly interacting superfluid Fermi gas. Through numerical simulations with time-dependent density-functional theory and the technique of the interaction quench, we verify the previous theoretical predictions on the mode's frequency. Next, we demonstrate that the mode is dynamically unstable against external perturbation and qualitatively examine the emerging state after the mode decays. The post-decay state is characterized by spatial fluctuations of the order parameter and density at scales comparable to the superfluid coherence length scale. We identify similarities with FFLO states, which become more prominent at higher dimensionalities and nonzero spin imbalances.
\end{abstract}

\maketitle

\section{Introduction}
\label{sec:intro}
Spontaneous breaking of continuous symmetries inevitably leads to the appearance of associated collective modes of a physical system. In the case of U(1) symmetry these are either gapless Goldstone mode, associated with oscillation of the phase of the order parameter, or the Higgs mode, which is associated with oscillations of the amplitude (for this reason, it is also called the amplitude mode). 
The Mexican-hat-like potential is typically drawn to visualize these modes: the Goldstone mode corresponds to motion around the hat in the minimum energy valley. In contrast, the Higgs mode is related to oscillations around the minimum in the perpendicular direction and consequently has much higher energy. 
Many studies related to the Higgs mode have been conducted in condensed matter (see eg.~\cite{Pekker2015} for review). In recent years, advances in cooling techniques also allowed investigating this phenomenon in highly controllable environments of ultracold atoms~\cite{Endres2012,Behrle_2018}, as well as developments in THz spectroscopy for superconductors~\cite{matsunaga2012nonequilibrium, matsunaga2013, matsunaga2014light, PhysRevLett.120.117001, chu2020phase, shimano2020}.
Simultaneously, many theoretical considerations have been presented concerning the Higgs mode in ultracold atomic systems: properties of small and large amplitude oscillations~\cite{Bulgac2009,PhysRevA.93.033641,PhysRevA.73.033614,PhysRevA.91.033628,collado2}, impact of the trapping potential on the mode characteristics~\cite{Tokimoto2017,PhysRevA.90.023621,Tokimoto2019}, angle-resolved photoemission spectroscopy (ARPES)~\cite{PhysRevB.92.224517, schwarz2020classification, PhysRevB.101.224510} and various other ways to induce collective modes~\cite{krull2016coupling, PhysRevB.96.184518, PhysRevB.95.104503, PhysRevB.104.L140503}.
Moreover, many analytical results have been derived in the weak-coupling limit~\cite{Barankov2004,barankov2006synchronization,PhysRevLett.96.097005,PhysRevA.91.033628,zhou}. The most well-known property is that the frequency of small-amplitude oscillations is related to the equilibrium value of the pairing gap by $\hbar\Omega=2\Delta$. Through this, one can deduce the pairing gap by measuring the Higgs mode frequency~\cite{Behrle_2018,Scott2012}.

The most widely discussed method of generation of the amplitude mode is through the interaction  quench~\cite{PhysRevA.91.033628,PhysRevB.99.014517,PhysRevA.100.013604,PhysRevResearch.3.023205,PhysRevLett.94.170402,PhysRevA.73.013609,PhysRevLett.105.135701,collado1}. This idea exploits the fact that the equilibrium value of the pairing field (order parameter) depends on the interaction strength, which is typically characterized by the dimensionless value $a\kF$, where $a$ is the scattering length and $\kF=(3\pi^2 n)^{1/3}$ is the Fermi wave-vector for a gas of density $n$. 
Thus, by preparing the system in the ground state for a selected initial interaction strength $a^{(\textrm{initial})}\kF$ and then changing the interaction regime rapidly to the new value $a^{(\textrm{final})}\kF$, one can induce oscillations of the order parameter within the range 
specified by $[\Delta(a^{(\textrm{initial})}\kF),\Delta(a^{(\textrm{final})}\kF)]$. Theoretical studies of such scenarios have been performed by use of the Bogoliubov de-Gennes (mean-field) method, which is justified for weakly interacting systems with attractive interaction: $|a\kF|\ll 1$ and $a<0$. On the other hand, Fermi superfluids produced in laboratories are typically strongly interacting $|a\kF|\gg 1$.

A separate aspect concerns the stability of the amplitude (Higgs) mode. In the case of weak couplings and small amplitude oscillations, by utilizing linear response theory,  Dzero, Yuzbashyan, and Altshuler demonstrated that the mode is unstable~\cite{Dzero2009CooperPT}. The mode decays into a new state with spatially nonuniform order parameter. Numerical simulations performed for the attractive  Fermi-Hubbard model~\cite{PhysRevB.99.035162,PhysRevB.99.014517} 
indicated that the mode is indeed dynamically unstable, which was consistent
with the theoretical predictions.

The purpose of the presented article is twofold. First, we examine the stability of the Higgs mode in strongly interacting Fermi gases -- the systems that are of particular experimental interest. By means of time-dependent density functional theory, we simulate the whole process: from the generation of the Higgs mode via the interaction quench, through the decay dynamics until the equilibration of the final (post-decay) state. 
Second, we investigate the properties of the post-decay state. We show that it is characterized by spatially inhomogeneous order parameter, bearing similarities to Larkin-Ovchinnikov (LO)~\cite{lo} or Fulde-Ferrel (FF)~\cite{ff} type oscillations. Upon introducing another degree of freedom to the problem in the form of spin imbalance, the post-decay state indeed acquires properties as predicted for spin-imbalanced systems ~\cite{arxiv.2211.01055,PhysRevA.100.033613,T_zemen_2020,PhysRevA.104.033304}, although the final state corresponds to an excited state. 

\section{Method and framework}
\label{sec:tddft+sldae}

We employ the Density Functional Theory (DFT) formalism to study the properties of the amplitude mode. Many variants of DFT methods exist; here we utilize the one known as Superfluid Local Density Approximation (SLDA), 
which has been specifically designed to simulate fermionic superfluid systems where interparticle interactions have short range~\cite{Bulgac2007,Bulgac2012}. It is a microscopic theory where the system is described in terms of fermionic quasi-particles, defined through Bogoliubov amplitudes: $\varphi_{n \sigma}(\bm{r},t)=[u_{n \sigma}(\bm{r},t), v_{n -\sigma}(\bm{r},t)]^T$, where $\sigma=\pm$ indicates spin projection.
Precisely, $v_{n \sigma}$ and $u_{n \sigma}$ stand for amplitude probabilities for $n$-th state to be occupied by a particle and a hole with spin projection $\sigma$. Formally, the equations of motion have the same structure as Bogoliubov-de Gennes (BdG), however they are obtained as a result of energy minimization expressed as a functional of several densities characterizing fermionic superfluid:
\begin{equation}
E = \int \mathcal{E}(n_\sigma,\tau_\sigma,\bm{j}_\sigma,\nu_\sigma)d\bm{r}.
\end{equation}
The energy density $\mathcal{E}$ depends on the following densities (we skip position and time dependence for brevity):
\begin{itemize}
 \item normal densities
 \begin{equation}
 n_{\sigma} = \sum_{|E_{n \sigma}|<E_c}\abs{v_{n \sigma}}^2 f_{\beta}(-E_{n \sigma}),
 \label{eqn:dens-n}
 \end{equation}
 \item kinetic densities
 \begin{equation}
 \tau_{\sigma} = \sum_{|E_{n \sigma}|<E_c}\abs{\nabla v_{n \sigma}}^2 f_{\beta}(-E_{n \sigma}), 
 \label{eqn:dens-tau}
 \end{equation}
 \item current densities
 \begin{equation}
 \bm{j}_{\sigma} = \sum_{|E_{n \sigma}|<E_c}\textrm{Im}[v_{n \sigma}\nabla v_{n \sigma}^{*}] \,f_{\beta}(-E_{n \sigma}),
 \label{eqn:dens-j}
 \end{equation}
 \item anomalous density
  \begin{equation}
  \nu_{+-} = \nu = \sum_{|E_{n \sigma}|<E_c} u_{n -}v_{n +}^{*}
  \frac{ f_{\beta}(-E_{n +} )-f_{\beta}(E_{n -})}{2}.
  \label{eqn:dens-nu}
  \end{equation}
\end{itemize}
The Fermi-Dirac distribution function $f_{\beta}(E) = 1/[1 + \exp \left ( \beta E \right )]$ accounts for temperature $T$ ($\beta^{-1}=k_B T$), although in this work we mainly focus on the $T\rightarrow 0$ limit. The densities are constructed from the quasiparticle amplitudes up to a certain energy cut-off $E_c$. This procedure accounts for regularization of divergences appearing due to short-range interactions~\cite{Bulgac2002a}. The minimization yields static equations for the quasi-particle amplitudes  (hereafter we use units where $m=\hbar=1$)
\begin{equation}
\begin{pmatrix}h_{\sigma}-\mu_{\sigma}  &  \sigma(\Delta + \delta\Delta_{\textrm{ext}}) \\ \sigma(\Delta^*+ \delta\Delta^*_{\textrm{ext}}) &  -h^*_{-\sigma}+\mu_{-\sigma}\end{pmatrix} \begin{pmatrix}u_{n \sigma} \\ v_{n -\sigma}\end{pmatrix}= E_{n \sigma}\begin{pmatrix}u_{n  \sigma} \\ v_{n -\sigma}\end{pmatrix}.
\label{eqn:HpsiEpsi}
\end{equation}
Note that there are two sets of equations for $\sigma=\pm$.
The time-dependent variant is obtained by replacing $E_{n\sigma}\rightarrow i\frac{\partial}{\partial t}$. 
Chemical potentials, denoted as $\mu_\sigma$, are used to control particle numbers $N_\sigma = \int n_\sigma\,(\bm{r}) d\bm{r}$ associated with a given spin component. \respadd{The time-dependent formulation conserves the total number of particles of each species $N_\sigma(t)=\textrm{const}$. }
The single particle Hamiltonian $h_\sigma$ and the pairing potential $\Delta$ are defined through functional derivatives of the energy:
\begin{equation}
h_{\sigma} = -\bm{\nabla}\frac{\delta E}{\delta \tau_\sigma }\bm{\nabla} + \frac{\delta E}{\delta n_\sigma}-\dfrac{i}{2}\left\lbrace \frac{\delta E}{\delta \bm{j}_\sigma},\bm{\nabla} \right\rbrace,\quad \Delta=-\frac{\delta E}{\delta \nu^*}.
\end{equation}
The pairing potential is a complex field $\Delta(\bm{r},t)$ that serves as the order parameter. In addition we have added an external pairing potential $\delta \Delta_{\textrm{ext}}$ that will be used to generate a perturbation when studying the stability of the amplitude mode. Due to symmetry of Eqs. (\ref{eqn:HpsiEpsi}) one needs effectively a solution for either $\sigma=+$ or $\sigma=-$. The other one can be obtained through symmetry transformation (see eg. \cite{PhysRevA.106.033322}).

The energy functional defines the system. In almost all cases devoted to the studies of the Higgs mode in Fermi systems, the mean-field Bogoliubov-de Gennes equations were used. They are valid for weakly interacting Fermi systems, and upon specific choice of the energy density functional
\begin{equation}
\mathcal{E} = \sum_{\sigma=\pm}\frac{\tau_\sigma}{2}+4\pi a |\nu|^2
\end{equation}
the formalism presented here becomes identical to the BdG. The single particle 
Hamiltonian and the pairing potential are then of the form:
\begin{equation}
h_{\sigma} = -\frac{1}{2}\bm{\nabla}^2,\quad \Delta=-4\pi a \nu.
\end{equation}
The advantage of a description based on DFT is that it allows to incorporate so-called ``beyond mean-field'' effects, while keeping the numerical complexity at a level comparable to the mean-field methods. Here we use the functional known as SLDAE~\cite{Boulet2022}, which has the generic form
\begin{equation}
\mathcal{E} = \sum_{\sigma=\pm}  \frac{A(a\kF)}{2}\left(\tau_\sigma-\frac{\bm{j}^2_\sigma}{n_\sigma}\right)
	        + \frac{3}{5}B(a\kF)\,n\,\eF
	        + \frac{C(a\kF) }{n^{1/3}} \abs{\nu}^2 + \sum_{\sigma=\pm} \frac{\bm{j}^2_\sigma}{2n_\sigma},
\end{equation}
where $n=n_{+} + n_{-}$ and $\eF=\frac{\kF^2}{2}=\frac{1}{2}[3\pi^2n]^{2/3}$ is the associated Fermi energy. The dimensionless coupling functions $A$, $B$ and $C$ are constructed in such a way to ensure the correct reproduction of thermodynamics quantities for a uniform system (equation of state, chemical potential, pairing gap, effective mass) over the entire interaction regime from weak (BCS regime) to strong (unitary Fermi gas regime). Their explicit forms
are given in paper~\cite{Boulet2022}. The SLDAE functional has been designed to provide accurate results for spin-symmetric systems ($N_{+}=N_{-}$), however it can be extended to spin-imbalanced systems as well. In this paper we will show a few selected results for spin-imbalanced systems ($N_{+}\neq N_{-}$). Since we are interested in qualitative aspects for such cases, we use the functional without further adjustments (in the same spirit as it is usually done with BdG approach). 
Note that further optimizations of the functional towards spin-imbalanced systems can be performed in a similar way as in Refs.~\cite{Bulgac2012,PhysRevLett.101.215301}, but from the perspective of this work it is not needed.

A quantity of special importance in the context of this work is the order parameter $\Delta(\bm{r},t)$. It is a complex function, and both attributes (magnitude and phase) carry physical information. The amplitude/Higgs mode corresponds to uniform oscillations of the absolute value across the whole system. Thus, we consider a uniform system (no external trapping potential) in which we induce the mode through the interaction quench method (see next section for details). 
At the beginning of each simulation the density distributions~(\ref{eqn:dens-n})-(\ref{eqn:dens-nu}) are uniform and may depend on time only.
In order to study the stability, we solve the equations of motion on spatial grid of size $N_x \times N_y \times N_z$ with lattice spacing $dx=dy=dz=1$ (definition of the length unit). The lattice spacing sets a natural cut-off energy scale $E_c\approx \frac{k_c^2}{2}=\frac{\pi^2}{2}$. Most of the calculations will be presented for the constrained case, where we impose translation symmetries along $y$ and $z$ directions. 
It implies that quasiparticle wave-functions have a generic structure that can be written as
\begin{equation}
\psi_n(\bm{r},t) \rightarrow \psi_n(x,t)\frac{1}{\sqrt{N_y}}e^{ik_y y}\frac{1}{\sqrt{N_z}}e^{ik_z z},\label{eqn:quasi-1d}
\end{equation}
where $k_y$ and $k_z$ take discrete values of multiples of $2\pi/N_y$ and $2\pi/N_z$, respectively. Under this assumption during the evolution (breaking of translation symmetry), the densities and the order parameter can acquire dependence along the $x$ direction only. To test the robustness of conclusions obtained from quasi-1D simulations, we have also performed additional simulations for quasi-2D case:
\begin{equation}
\psi_n(\bm{r},t) \rightarrow \psi_n(x,y,t)\frac{1}{\sqrt{N_z}}e^{ik_z z},\label{eqn:quasi-2d}
\end{equation}
as well as fully unconstrained simulations in 3D. We specifically focus on the large amplitude mode induced in the strongly interacting regime ($|a\kF|\rightarrow\infty$), as the stability for this case has not been discussed in the literature to date.
For the computation we use W-SLDA Toolkit \cite{PhysRevLett.120.253002,PhysRevLett.112.025301,WSLDAToolkit}. All technical parameters, needed to restore the calculations presented below are included into reproducibility packs~\cite{SM}. They contain also detailed information about the numerical setup and computation process.

\begin{figure}[t]
\centering
\includegraphics[width=\textwidth]{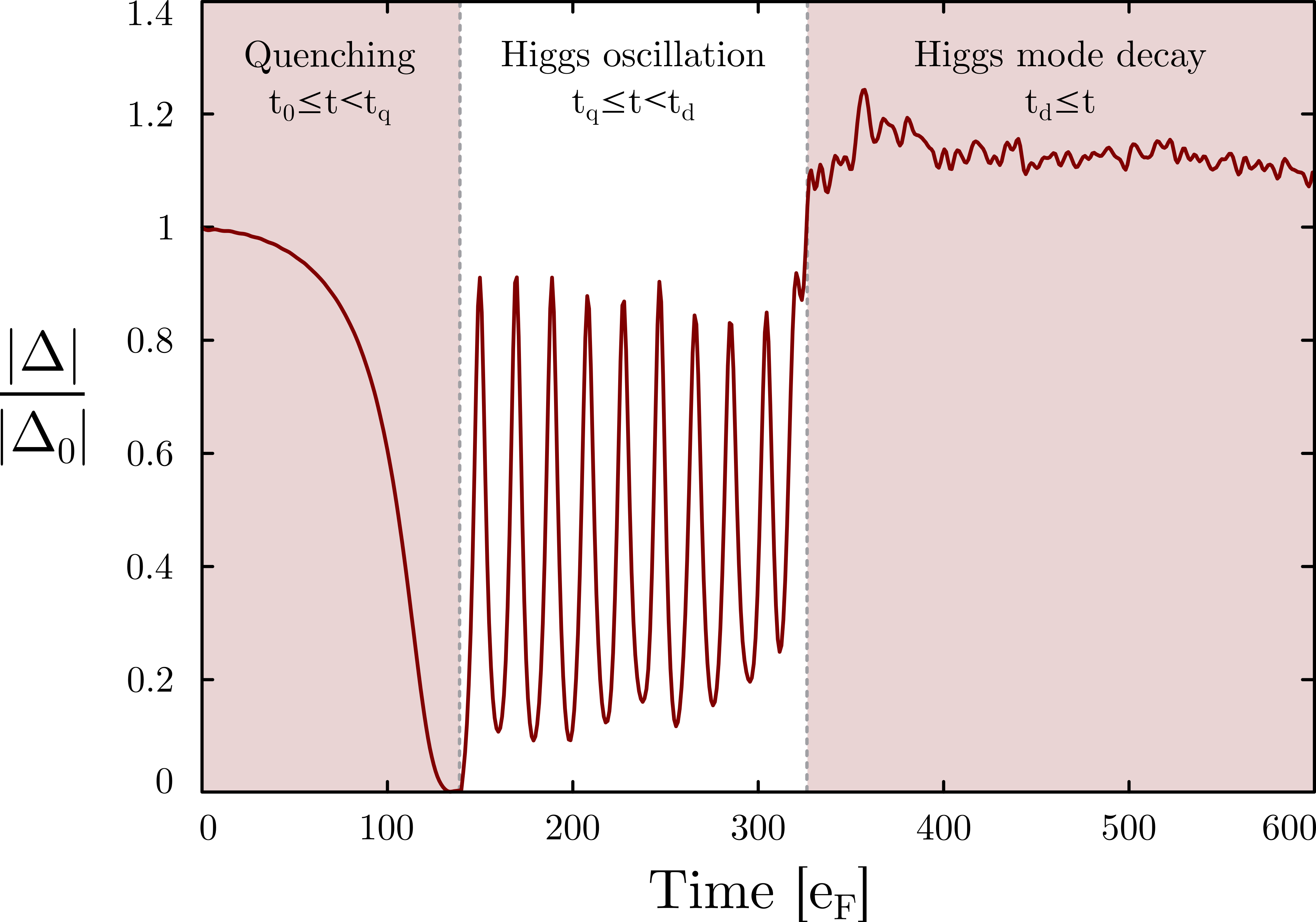}
\caption{The numerical experiment investigated in this work. The Higgs mode is induced in the uniform system by quenching the interaction strength in the time interval $t_0(=0)<t \leq t_q$. The mode persists for some time ($t_q<t \leq t_d$), after which it decays into a nonuniform state ($t>t_d$). The red line presents the absolute value of the pairing field for a selected point at the center of the lattice, $|\Delta(x=0, t)|$, normalized to the initial value. The presented case corresponds to a large amplitude Higgs mode.}
\label{fig:quench}
\end{figure}

\newpage
\section{Generation of the Higgs mode via the interaction quench}
\label{sec:hm-generation}

Rapid quenching of the interaction strength $a^{(\textrm{initial})}\kF\rightarrow a^{(\textrm{final})}\kF$ is the standard method of inducing the Higgs mode~\cite{Barankov2004,Bulgac2009,Scott2012,Tokimoto2017,Tokimoto2019,arxiv.2210.09829}. 
After the generation of the Higgs mode, it is interesting to test its stability with respect to random perturbation. The perturbation is realized by adding the weak external potential $\delta \Delta_{\textrm{ext}}(\bm{r},t)$. Consequently, in our simulations the emerging time evolution can be divided into 4 parts. 

\begin{description}
    \item[Initial configuration] ($-\infty < t \leq t_0$)\par
        The homogeneous gas characterized by $s$-wave scattering length $a_0$ and (normal) density $n_0 = \kF^3 / (3\pi^2)$ is in its ground state. We consider spin-symmetric simulations ($N_{\uparrow}=N_{\downarrow}$), and only in some cases we start simulations from states corresponding to spin-imbalanced ($N_{\uparrow}\neq N_{\downarrow}$) systems. 

    \item[Quenching of interaction] ($t_0< t \leq t_q$)\par
        The initial state is driven adiabatically towards the new interaction strength $a_0\kF\rightarrow a_1\kF$ in the time period $t_0<t\leq t_1$, and then quasi-instantaneously driven back to the initial value $a_1\kF\rightarrow a_0\kF$ in the time interval $t_1<t\leq t_q$, where $(t_1-t_0) \gg (t_q-t_1)$. Explicitly, the time-dependent $s$-wave scattering length is parametrized as follows:
\begin{align}
    a(t) = 
    \begin{cases}
    a_0 + \dfrac{a_1-a_0}{2}[1 - \cos\omega_1(t-t_0)],
    & t_0 < t \le t_1
    \\
    a_1 + \dfrac{a_0-a_1}{2}[1 - \cos\omega_2(t-t_1)],
    & t_1 < t \le t_q
    \end{cases}
\end{align} 
such that $a(t)$ and the first derivative $\partial_t a(t)$ are continuous, leading to $t_1 - t_0 = {\pi}/{\omega_1}$ and $t_q - t_1 = {\pi}/{\omega_2}$. Using this parametrization, the limit $\omega_1 \ll \eF$ corresponds to an adiabatic process, i.e. a small energy transfer to the system. On the other hand $\omega_2 \gtrsim \eF$ generates the rapid quench. 
    
    \item[Higgs oscillations] ($t_q < t \leq t_d$)\par
    The interaction quench induces the Higgs oscillations with wave vector $k=0$ (uniform oscillations) and frequency $\Omega$. Thus, observables do not exhibit position dependence, for example for the pairing field we have $\vert \Delta(x,t) \vert \rightarrow \vert\Delta(t)\vert$. In order to investigate the stability of this mode, we add a weak perturbation to the system $\delta \Delta_{\textrm{ext}}(\bm{r},t)$ once the oscillations have developed. We have tested the stability of the Higgs mode with respect to two types of perturbation: local in momentum space and spatially localized. The explicit forms of perturbations read as: 
    \begin{enumerate}
    \item $\delta\Delta_\text{ext}^{(1)}(x,t) = \epsilon \exp(ikx) \exp(-(t-t_p)^2/2\sigma^2)$ (induced decay by spatial modulation)
    \item $\delta\Delta_\text{ext}^{(2)}(x,t) = \epsilon \delta(x) \exp(-(t-t_p)^2/2\sigma^2)$ (induced decay by Dirac-delta perturbation)
    \end{enumerate}
    where $\epsilon\ll\eF$ (weak perturbation) and $\sigma\eF\lesssim 1$ \respadd{(applied at much shorter times scale as the expected time scale of the Higgs oscillation)}. In numerical realization the Dirac function $\delta(x)$ is approximated by narrow Gaussian function. 

    \item[Higgs mode decay] ($t_d < t$)\par
    After some time $t_d$ we find that the Higgs mode decays and an inhomogeneous phase emerges. Namely, the observables depend now on position and time, for example for the quasi-1D calculations: $n(x,t) \neq n_0$ and $\Delta(x,t) \neq \Delta(t)$.
\end{description}
In Fig.~\ref{fig:quench} we present an example time evolution of the absolute value of the order parameter for selected point in space $|\Delta(x=0, t)|$. The described stages of the numerical experiment can be easily identified. 

As it will be shown in this paper, the Higgs mode is dynamically unstable, i.e small perturbations amplify and destroy the mode. We note that even in the case of $\delta\Delta_\text{ext}(x,t) = 0$ (no external perturbation) the mode decays spontaneously after some time $t_{\textrm{max}}$. In such case, the mode destabilizes due to numerical noise, since we integrate the equations of motion numerically with some precision. When measuring the lifetime of the Higgs mode, defined as the time from the weak perturbation to the decay, we limit our considerations to time intervals $t<t_{\textrm{max}}$ to rule out complications related to imperfections of the numerical integration. 
We recognize that in the case of studies of the dynamical stability of physical phenomena by means of computational physics, code quality is an important issue. We note that the accuracy of time integrator implemented within W-SLDA, which is multistep Adams-Bashforth-Moulton of 5th order, has been tested in work~\cite{PhysRevC.105.044601}. 
Namely, the dependence of the results with respect to the perturbations of the initial states, for cases where no dynamical instability is expected, was tested. Moreover, the W-SLDA Toolkit has been applied to a variety of problems. Results were documented in papers~\cite{arxiv.2207.06059,PhysRevLett.130.043001,PhysRevA.104.033304,PhysRevLett.120.253002,PhysRevA.100.033613,PhysRevA.105.013304,PhysRevA.103.L051302}. We have confirmed stability of the results with respect to size of the integration time step $dt$ (all presented here results were obtained with $dt=0.0025\eF^{-1}$). \respadd{We have also checked the stability of the results with respect to the lattice size. For the quasi-1D calculations (see sections~\ref{sec:hm1d} and \ref{sec:postdecay}), we have checked that conclusions are stable if we vary lattice size from 256 to 1024.}
Thus we regard the code as well tested, and the observed instability in the simulations is expected to be due to intrinsic properties of the studied problem, not due to artifacts generated by the numerical implementation. 

\section{Properties of the Higgs mode oscillations}
\label{sec:propHM}

We start by examining the properties of the amplitude mode. The aim of this section is to demonstrate that the interaction quench method induces the Higgs mode, having the well known properties. The mode is typically regarded as as oscillating system in the effective potential in the form of a Mexican hat:

\begin{equation}
V(\Delta) = -\frac{1}{2}\mu^2\abs{\Delta}^2 + \frac{1}{4}\epsilon\abs{\Delta}^4.
\end{equation}
The solutions for a classical particle moving in such potential are well known~\cite{Pekker2015,arefeva2012rolling}. In the case of small amplitude oscillations around the minimum of the potential $\Delta_0=\pm \mu/\sqrt{\epsilon}$ we have:
\begin{equation}
\Delta(t) = \Delta_0 + A(t)\sin(\omega t+\phi),
\end{equation}
where $\omega$ is the oscillator's frequency, $\phi$ is its phase, and  $A(t)$ is the amplitude, which in general may depend on time~\cite{PhysRevA.91.033628,barankov2006synchronization}. Moreover, for the small amplitude Higgs mode, the oscillation frequency is related to the equilibrium value of the pairing gap $\hbar\omega=2\Delta_0$. The oscillation of arbitrary amplitude are expressed by delta amplitude Jacobi elliptic function $\textrm{dn}$:
\begin{equation}
    \label{eq:dn}
    |\Delta(t)| = |\Delta(t_0)| \textrm{dn}(\Omega t+\phi,k),
\end{equation}
where $\Delta(t_0)=\Delta_0=0.4425e_F$ and the elliptic modulus parameter $k$ is related to maximum and minimum values of the paring field during the oscillations:

\begin{equation}
    \label{eq:k-dn}
    k \approx 1-\left(\frac{\min[\Delta(t)]}{\max[\Delta(t)]}\right)^2
\end{equation}

\begin{figure}
\centering
\includegraphics[width=\textwidth]{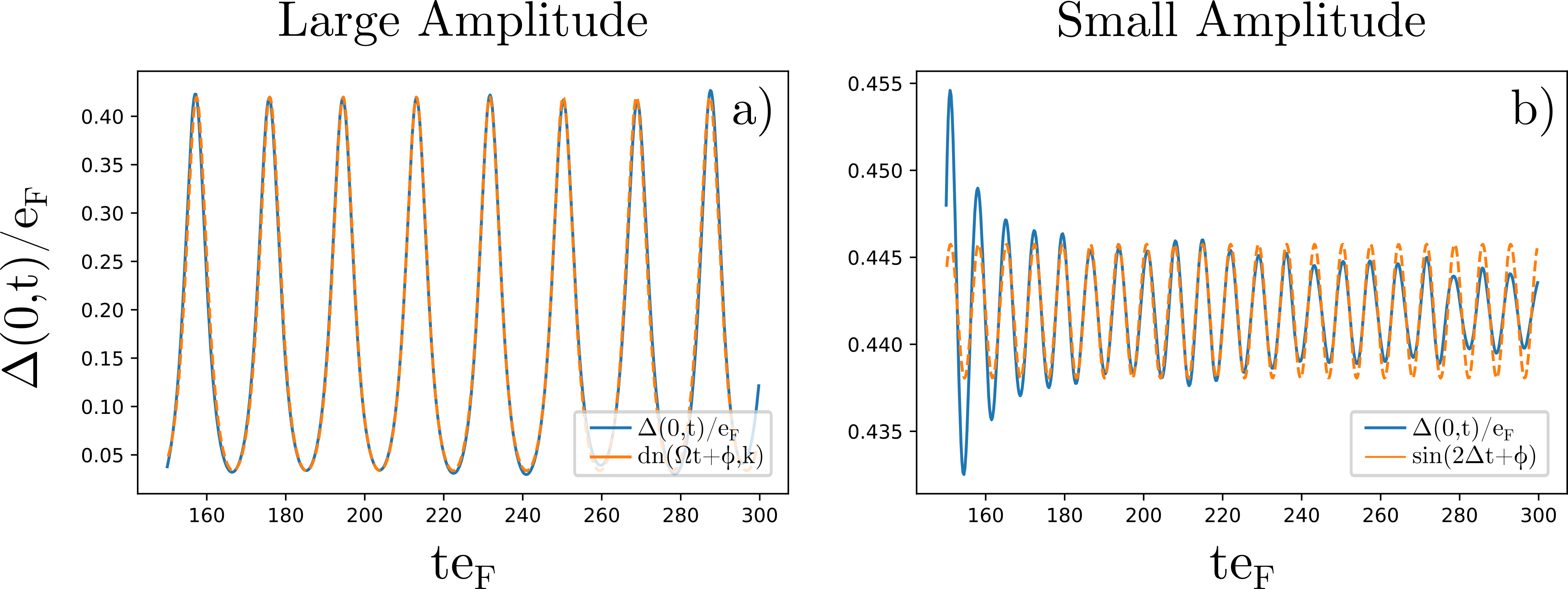}
\caption{Time evolution of the pairing field (order parameter) magnitude in the center of the simulation domain $\Delta(x=0,t)$ for times corresponding to the Higgs oscillation regime. (a)~Large amplitude Higgs oscillations (blue line), induced by rapidly quenching the interaction from $a_1 k_F=-0.1$ to $a_0 k_F=-10$. They are well reproduced by elliptic function~(\ref{eq:dn}) with $\Omega=0.95\Delta_0$ and $k=0.9935$ (orange line). (b)~Small amplitude Higgs mode (blue line), induced by rapidly quenching the interaction from $a_1 k_F=-5$ to $a_0 k_F=-10$. The oscillation frequency is $\omega=2\Delta_0$ (orange line).}
\label{fig:omega}
\end{figure}

In Fig.~\ref{fig:omega} we compare the numerically extracted time dependence of the pairing field $\Delta(t)$ after the interaction quench with analytic predictions. The time-dependent variant of Eq.~(\ref{eqn:HpsiEpsi}) has been solved numerically on a lattice of size $256 \times 32 \times 32$ within quasi-1D geometry, see Eq.~(\ref{eqn:quasi-1d}).
In both cases, small and large amplitude oscillations, we observe a close match with analytic predictions. 
In particular, the well known property of the small amplitude Higgs mode, $\omega=2\Delta_0$, has been reproduced with very good accuracy. The large amplitude mode oscillates with lower frequency $\Omega\approx \Delta_0$ and the parameter $k=0.9935$ provided by the fit also agrees well with the expected value $k=0.9952$ obtained by applying Eq. \ref{eq:k-dn}. These findings are fully consistent with previous works, such as~\cite{PhysRevA.91.033628,barankov2006synchronization,Bulgac2009}.

\section{Lifetime of the Higgs mode}
\label{sec:hm1d}

\begin{figure}[t]
\centering
\includegraphics[width=\textwidth]{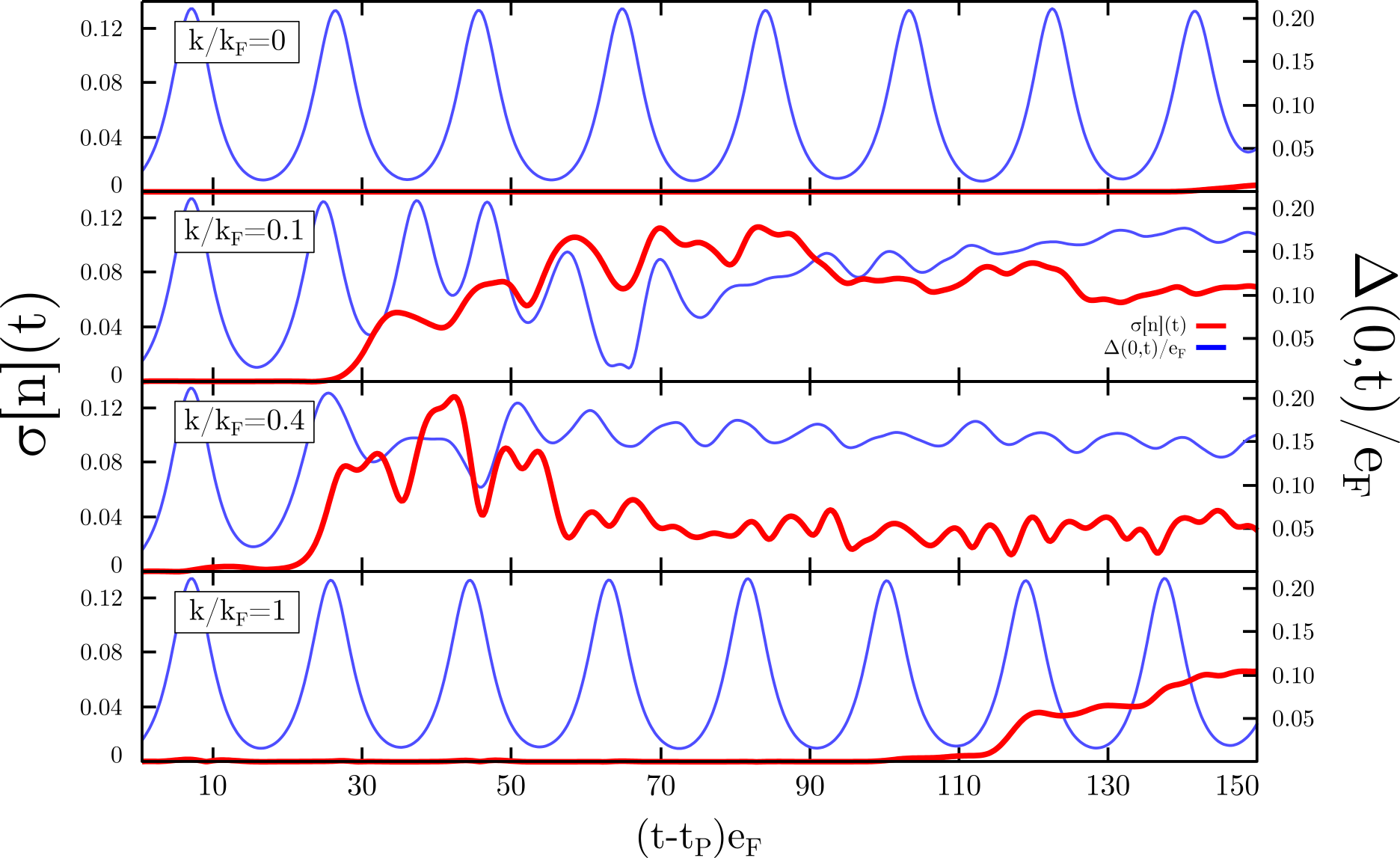}
\caption{Time evolution of the normalized standard deviation of density $\sigma[n](t)$ and amplitude of the order parameter in selected point in space $\Delta(0,t)$ after the applied perturbation in form $\Delta_\text{ext}^{(1)}(x,t) = \epsilon \exp(ikx) \exp \big(-\frac{(t-t_p)^2}{2\sigma^2} \big)$. In these simulations the large amplitude mode is induced by rapidly quenching the interaction from $a_1\kF=-0.1$ to $a_0\kF=-10$. The perturbation function parameters are $\epsilon/\eF=10^{-3}$ and $\sigma\eF=0.5$. Each panel corresponds to various wave-vectors $k$, indicated on the respective labels. Calculations were executed with quasi-1D geometry on a lattice of size $512 \times 32 \times 32$.}
\label{fig:decay_time}
\end{figure}

\begin{figure}[t]
\centering
\includegraphics[width=0.49\textwidth]{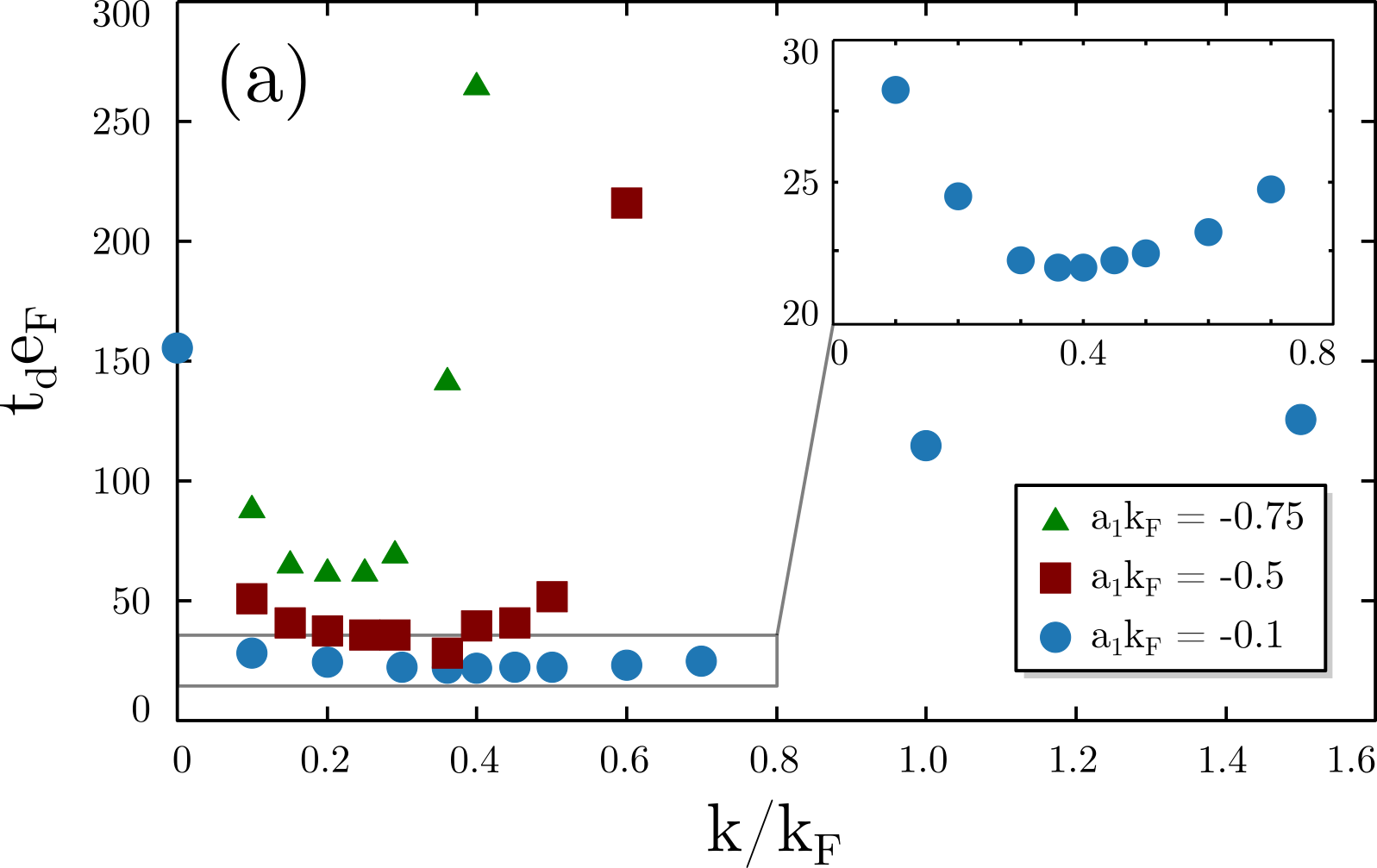}
\includegraphics[width=0.49\textwidth]{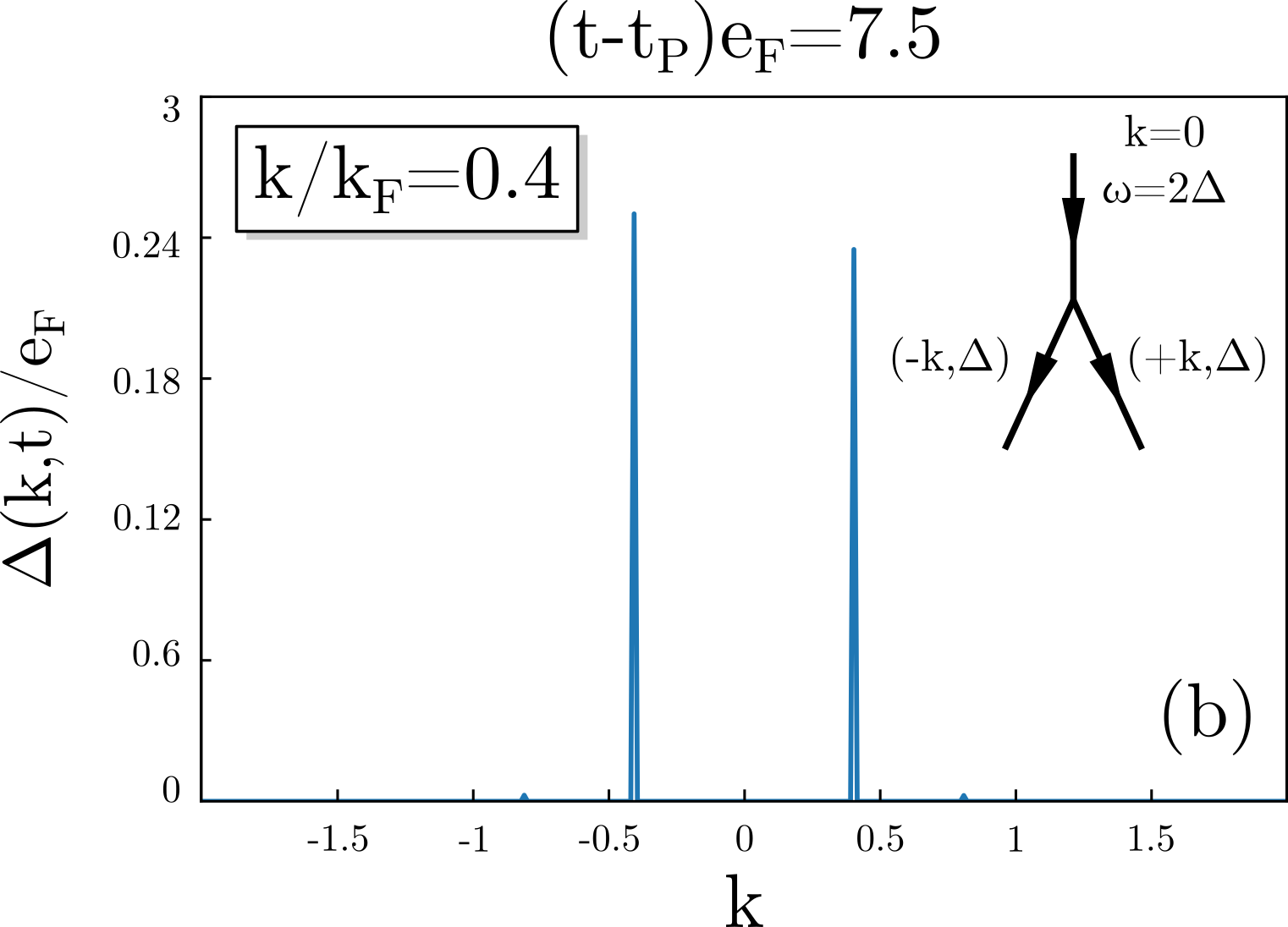}
\caption{Panel (a): Decay time as a function of the modulation wave vector $k/k_F$ for different stregths of the quench; values on the label indicate the values of $a_1\kF$ and $a_0\kF=-10$ is fixed. Panel (b): Fourier spectra of the order parameter just before the decay induced by the external modulation $k/k_F=0.4$ for the case with  $a_1\kF=-0.1$. Inset: The decay diagram for the small amplitude Higgs mode.}
\label{fig:decay_k}
\end{figure}

The Higgs mode is attributed to oscillations of the amplitude of the order parameter, while the density of the system remains unchanged. Indeed, when the mode is induced in a uniform system, $n(\bm{r}, t)$ stays constant. The appearance of inhomogenities in the density is associated with the decay of the Higgs mode. To quantify the magnitude of spatial variations of the density we compute:
\begin{equation} \label{eq:h-order-param}
    \sigma[n](t) = \frac{\sqrt{\langle n^2 \rangle(t) - \langle n \rangle^2(t)}}{\langle n \rangle(t = 0)},
    \quad \text{with} \quad 
    \langle n^k \rangle(t) = \frac{1}{V}\int  n^k(\bm{r},t)\,\mathrm{d}\bm{r}.
\end{equation}
The decay time $t_d$ is defined as the time for which the normalized standard deviation exceeds a threshold value $\sigma[n](t>t_d)>\gamma=0.01$. The dynamics of real system is always attributed by stochastic fluctuations (for example, due to thermal effects) that may amplify in time in case of unstable modes. In the numerical scenario, we introduce the weak external perturbation $\delta\Delta_\text{ext}$ and check if induced fluctuations amplify or decay. In the case of the mode studied here, we find that it is unstable. 
One can also induce fluctuations by adding weak external potential that couples to the density. The final qualitative result does not depend on the perturbation method. We define the lifetime of the Higgs mode as the time from the external perturbation $t_p$ to the decay $t_d$. In Fig.~\ref{fig:decay_time} we have shown induced decay by the perturbation $\delta\Delta_\text{ext}^{(1)}(x,t)$. We clearly observe a sensitivity of the decay time with respect to the spatial modulation wavelength $\lambda \sim 1/k$. 
The perturbation with $k=0$ (top panel) corresponds to preserving translational symmetry and therefore it does not have any impact on the dynamics of the system. The mode starts decaying due to the amplification of numerical errors only, after a time $t_{\textrm{max}}\eF\gtrsim 150$. 
On the other hand, perturbations corresponding to $k>0$ induce a decay whose onset time $t_d$ depends on the wavelength of the spatial modulation, see Fig.~\ref{fig:decay_k}(a). 
Clearly, there exist optimal value of $k$ (resonance) that trigger the instability almost immediately. \respadd{The location of the optimal value depends on the interaction strength quench. For tested cases it is located around $k/\kF\approx 0.2 - 0.4$.}
In Fig.~\ref{fig:decay_k}(b) we show the Fourier spectra of the order parameter after applying the perturbation, but before the decay. It is seen that the uniform Higgs mode ($k=0$) first decays into two modes with $+k_{d}$ and $-k_{d}$. The diagram representing schematically the decay of small amplitude mode is shown in inset of Fig.~\ref{fig:decay_k}(b).  

\begin{figure}[h]
\centering
\includegraphics[width=0.49\textwidth]{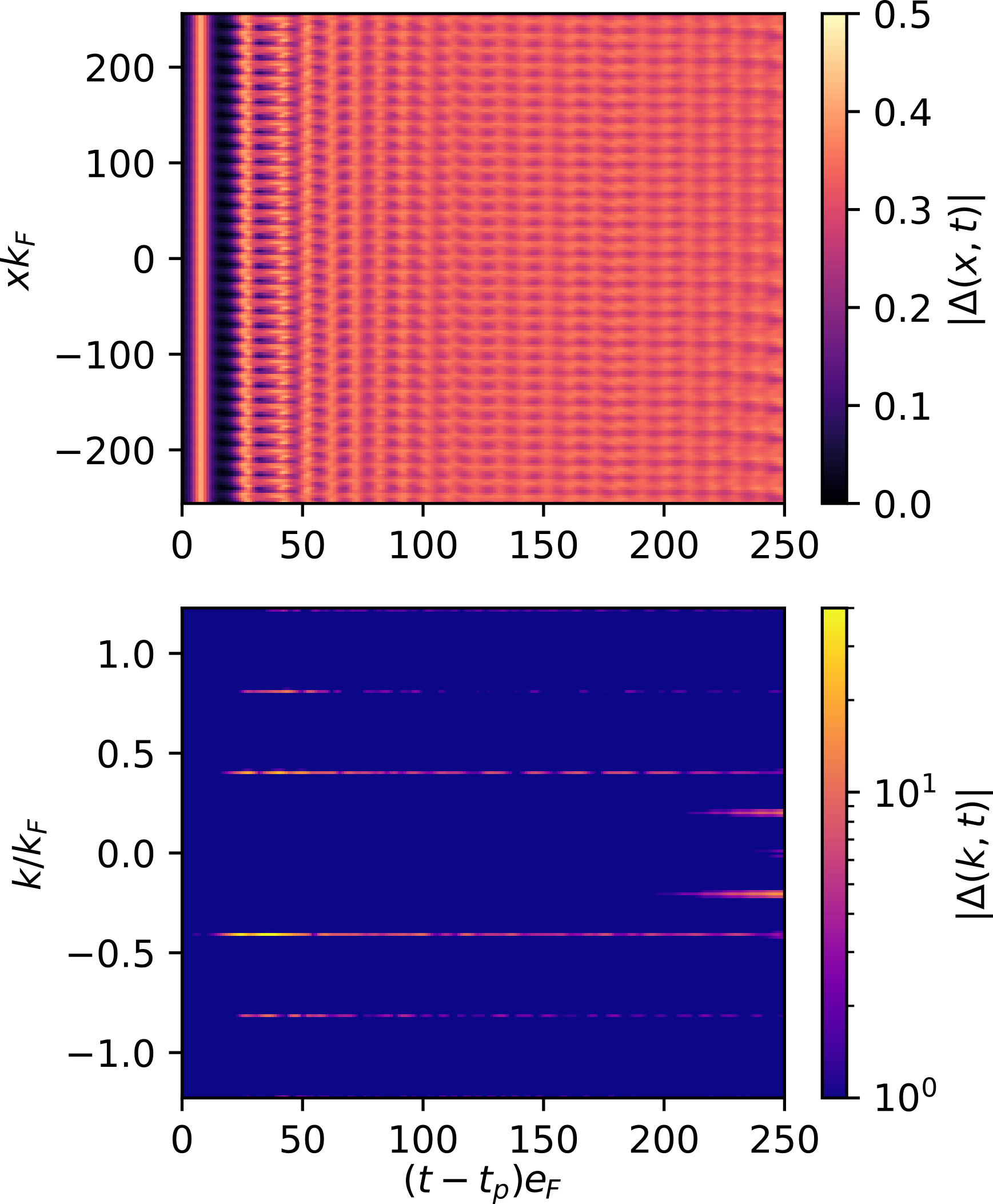} 
\includegraphics[width=0.49\textwidth]{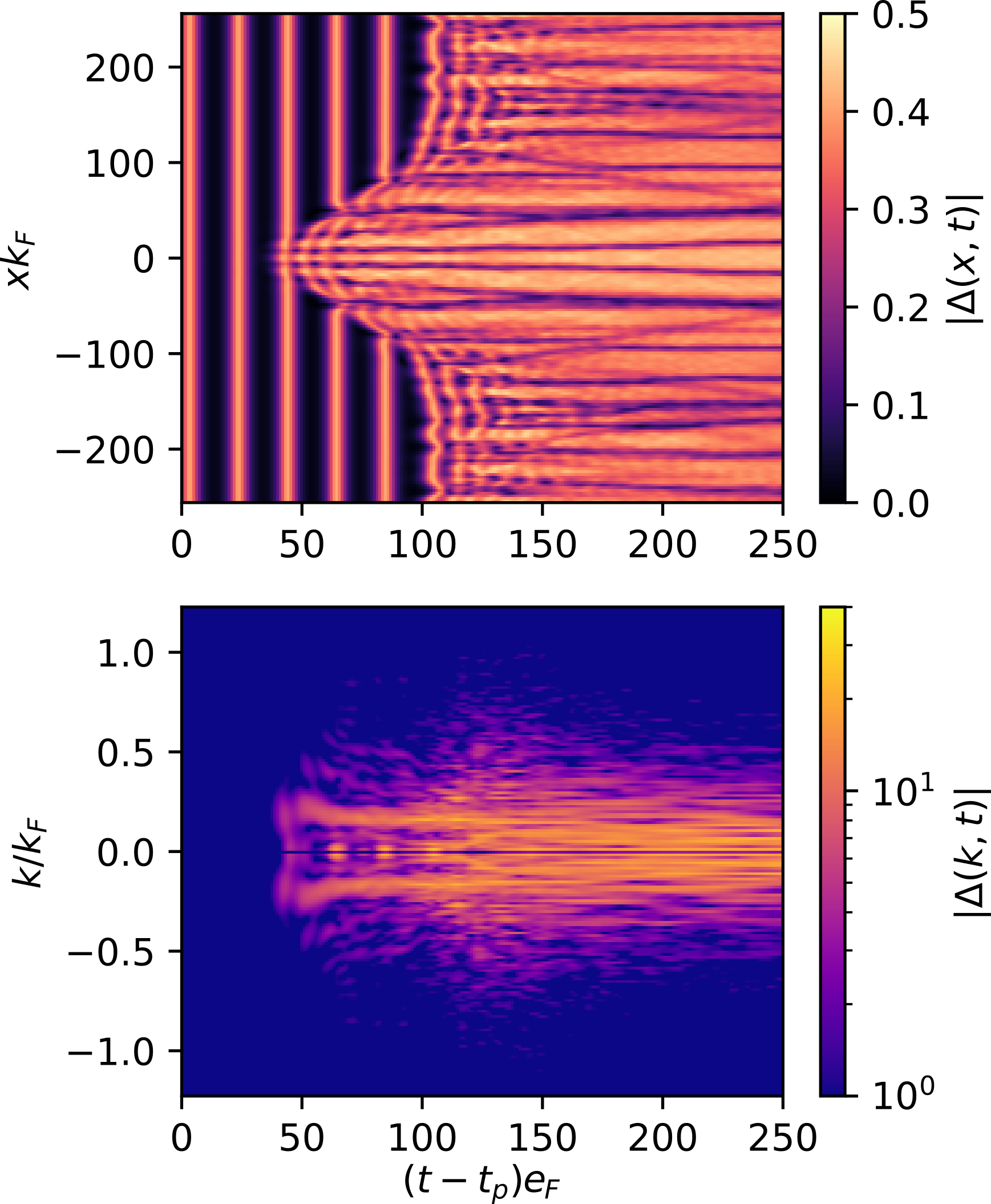}
\caption{Evolution of the order parameter in coordinate space (top row) and its Fourier spectra (bottom row) for perturbation $\delta\Delta_\text{ext}^{(1)}(x,t)$ with $k/\kF=0.4$ (left column) and $\delta\Delta_\text{ext}^{(2)}(x,t)$ (right column). The $\Delta(k=0,t)$ component was removed from the Fourier spectra for better visibility. The amplitude of the perturbation was $\epsilon/\eF=10^{-3}$ and evolution is shown only after the perturbation is applied. Maps generated by Matplotlib library~\cite{Hunter:2007}.}
\label{fig:decay_map}
\end{figure}

Instead of using a perturbation with well defined wave length $2\pi/k$, one can instead use a spatially localized perturbation such as $\delta\Delta^{(2)}_\text{ext}\sim \delta(x) = \int \frac{dk}{2\pi}e^{ikx}$. In numerical realization, we model the $\delta(x)$ function by a narrow Gaussian. Similarly, we observe that the amplitude mode decays and the state after the decay is nonuniform. In Fig.~\ref{fig:decay_map} we compare spatio-temporal evolution of the order parameter after applying different types of perturbation. 
The decay dynamics depend on the perturbation type; however there are gross properties that remain unchanged. Namely, in the Fourier spectra of the order parameter $|\Delta(k,t)|$ we observe dominant wave-vectors $\pm k_d$ into which the mode decays. These are typically in the range of $0.1\lesssim k_d/\kF\lesssim0.5$. It is comparable to the wave-vector associated with the coherence length scale $k_{\xi} = 1/\xi = \Delta/k_F$. Namely, for results presented in Figs~\ref{fig:decay_time}-\ref{fig:decay_map} the wave-vector associated with the coherence length reads $k_{\xi}/\kF\approx 0.22$, which translates into $0.5\lesssim k_d/k_{\xi}\lesssim2.3$. 

The results obtained for the $\delta$-kick need extra clarification (top right panel of Fig.~\ref{fig:decay_map}). In general, one can expect that the fluctuation induced by a localized perturbation should propagate with speed not exceeding the speed of sound.  Indeed in such case, we should see the perturbation only in the causality cone. It is visible only for times  $(t-t_p)e_F\lesssim75$. For later times one may conclude from the graph that the causality is violated. Such a conclusion is incorrect. The observed effect is the numerical artifact. Namely, we model $\delta(x)$ function by narrow gaussian, as the numerical scheme assumes that all derivatives are continuous and smooth. This implies that our perturbation is not localized in practice but affects the system in the whole spatial domain. 

Dzero, Yuzbashyan \& Altshuler~\cite{Dzero2009CooperPT} demonstrated that the small amplitude Higgs mode is unstable. Their result has been obtained within mean-field (BdG) treatment, which is valid for weakly interacting superfluid Fermi gases. In such case, the presence of dynamical instability was demonstrated analytically within the framework of the linear response theory. In the present study, we have released these constraints by considering large amplitude and strongly interacting regime and we have demonstrated the instability of the mode.
Namely, we have shown that similarly to the small amplitude regime, the mode decays predominantly into a state where the spatial modulation of the pairing field is given by $\sim 1/k_{d}$, i.e. $\Delta(x,t)=\Delta(t) + A(t)[e^{-ik_d x} + e^{ik_d x}]$, where $A(t)$ is the amplitude of the perturbation which grows rapidly in time.
Consequently, this implies that after the decay, the order parameter exhibits fluctuations of the type $\Delta(x) \sim \cos(k_d x)$. Indeed, in Fig.~\ref{fig:decay_map} (right panel) one can clearly see lines where $|\Delta(x,t)|=0$ (called nodal lines) for later times. 
These are lines where the order parameter changes sign. The spatial fluctuations of $\Delta(x) \sim \cos(k x)$ are typically regarded as a fingerprint of Larkin-Ovchinnikov (LO) state~\cite{lo}, if they emerge in spin-imbalanced system in a ground state. None of these requirements is satisfied in the cases discussed so far. However, it is interesting to note that the state emerging from the decay process shares similarities with the exotic type of superfluidity. This issue will be discussed in more detail in next section.

\section{Properties of the post-decay state}
\label{sec:postdecay}

\begin{figure*}[t!]  
\includegraphics[width=\textwidth]{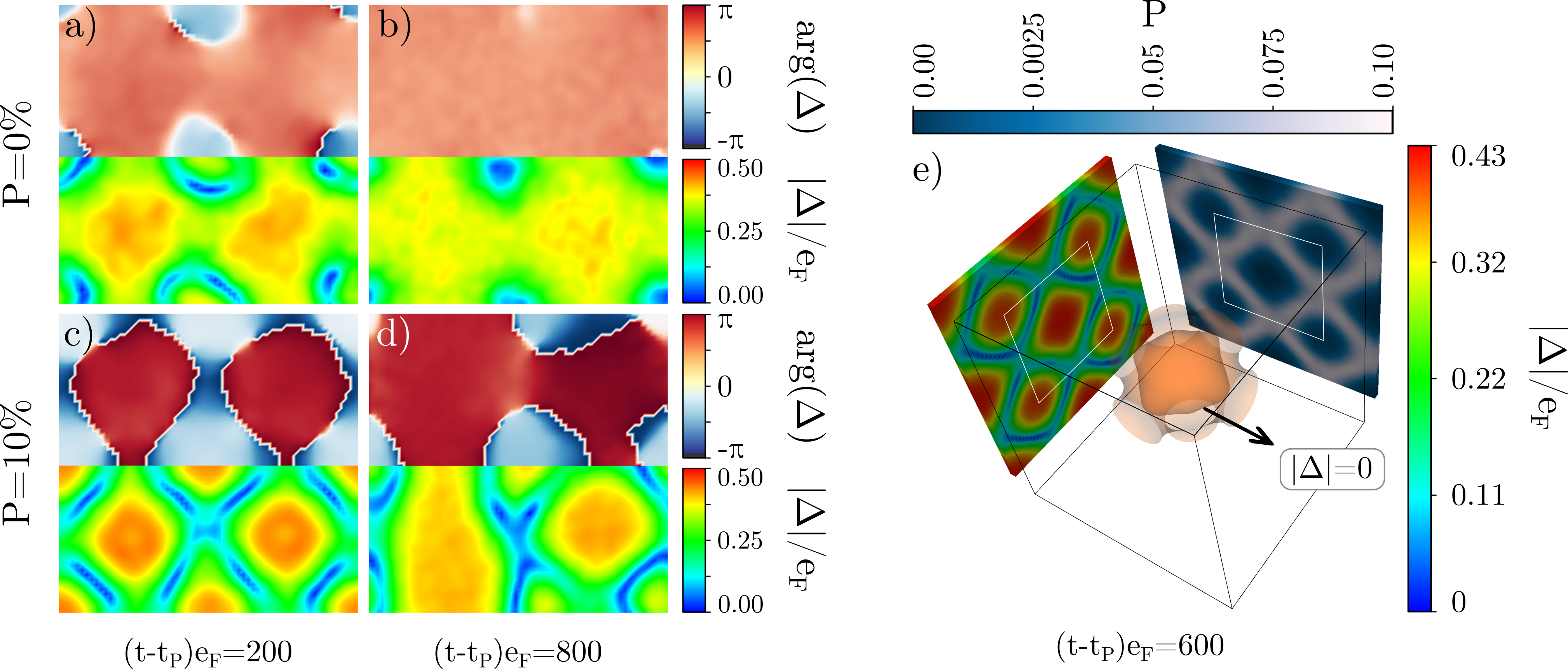}
    \caption{(a) and (b): 2D configurations for $P=\frac{N_\uparrow-N_\downarrow}{N_\uparrow+N_\downarrow}=0\%$, respectively after a short and long time from the perturbation is applied. 
    (c) and (d): 2D configurations for $P=10\%$.
    Panels on the bottom row (rainbow palette) show the absolute value of the order parameter, normalized to $e_F$; panels on the top row (heat palette) show the phase of the order parameter. 
    e) 3D configuration for $P=5\%$. Upper colorbar indicates local spin polarization $p(\bm{r})$ on the right panel; colorbar on the right shows the absolute value of the order parameter $|\Delta(\bm{r})|$, normalized to $e_F$. Inner inset shows the contour surface for which the order parameter is zero. The dimensionless size of the box is $L\kF=64$ in each direction. Maps and 3D view created by VisIt software~\cite{HPV:VisIt}}
    \label{fig:glass}
\end{figure*}

We now consider the properties of the state after the decay. To better visualize the typical characteristics of the post-decay state, we switch to a quasi-2D geometry, ansatz~(\ref{eqn:quasi-2d}). The calculations were executed on a $64\times 64 \times 16$ lattice. Qualitatively, we observe the same phenomena as reported above for quasi-1D cases: the mode decays to a state that is attributed by the non-homogeneous distribution of the order parameter. The structure of $\Delta(x,y)$ consists of regions where the phase changes by $\pi$, which are separated from each other by nodal lines defined as $|\Delta(x,y)|=0$.
The development of inhomogeneities of the pairing field is accompanied with spatial modulations of density distribution which are induced simultaneously. One needs to emphasize that it is a transient state, which evolves on much longer time scale (see Fig.~\ref{fig:glass}(a,b) with comparison at different times from the perturbation). For later times (larger time scale than reachable in numerical computation), we expect the system to thermalize again to the state with uniform density distribution and the order parameter. 

Interestingly, the transient state that emerges from the decay shares similarities to states as predicted for spin-imbalanced systems~\cite{PhysRevA.100.033613,T_zemen_2020,arxiv.2211.01055}. The main difference is that if the imbalance is present ($N_\uparrow \neq N_\downarrow$), the majority particles tend to accumulate in the nodal regions if such are developed in the system. 
It can be understood from the point of view of the system's energetics: not all particles can form Cooper pairs (superfluid component), and the energetically lowest price is paid if we store them close to the nodal lines where the pairing correlations vanish. The local spin polarization, defined as $p(\bm{r})=[n_\uparrow(\bm{r})-n_\downarrow(\bm{r})]/[n_\uparrow(\bm{r})+n_\downarrow(\bm{r})]$, acts as a stabilizer and converts the states consisting of nodal lines into meta-stable structures. Such structures, stabilized by the spin polarization, were recently the subject of studies where they were called ferrons~\cite{PhysRevA.100.033613} or ring solitons~\cite{babaev}. The size of these grains depend on the global population imbalance of the system $P=[N_\uparrow-N_\downarrow]/[N_\uparrow+N_\downarrow]$. For small imbalances $P\lesssim 10\%$ their sizes can by of the order $\sim 10\xi$, and decrease to $\sim\xi$ as the imbalance increases, and finally converting to well known LOFF state~\cite{arxiv.2211.01055}. 
Suppose we induce the Higgs mode in the spin-imbalanced system and let it decay, precisely in the same manner as we did for the spin-symmetric case.
Then, we indeed find that the post-decay state is stabilized by the spin-polarization, which at time scales larger then the time scale associated with the Higgs mode decay evolves towards state consisting of many spin-polarized impurities, see Fig.~\ref{fig:glass}(c,d). Note that just after the decay, the typical modulation wave-length for the order parameter (and also for the density) is close to the the wave-length
of the fastest decaying mode. Precisely, the average size of the structures we observed in Figs~\ref{fig:glass} (a) and (c) translate into wave-vectors $k/\kF=2\pi/\lambda\kF\approx 0.2$ for both $P=0\%$ and $P=10\%$ at the onset of the oscillation. 
Clearly, these values to the minimum location in Fig.~\ref{fig:decay_k}. Once the disordered structure is set by the decayed Higgs mode properties, it is driven to the new configuration  by  effects related to dynamics of the spin polarization. They operate on longer time scales, and in large time scales ($t\eF \sim 10^3$) the polarized system evolves towards to the known from past studies configurations~\cite{PhysRevA.100.033613,T_zemen_2020,arxiv.2211.01055}.

Finally, we demonstrate that the observations hold for fully 3D cases, as shown in Fig.~\ref{fig:glass}(e). These results were obtained for a lattice of size $64 \times 64 \times 64$. Similarly to the quasi-2D case, we introduce the spin-imbalance to the system in order to stabilize the final state. The post-decay state, at large times, consists of many bubble-like structures. Inside each bubble, the phase of the order parameter is shifted by $\pi$. The overall properties of the post-decay state remain fully consistent with results obtained for simplified geometries (quasi-1D and quasi-2D). The insensitivity of the results at the quantitative level with respect to the computation dimensionality demonstrates the robustness of derived conclusions.

\section{Conclusions}
We have studied the life cycle of an ultracold atomic gas after the interaction quench from weak to strong coupling. The quench applied to uniform system induces the large amplitude Higgs/amplitude mode. This mode turns out to be dynamically unstable and it decays to a state with spontaneously broken spatial symmetry. It is of a different type than discussed so far in literature~\cite{VolkovJETP,PhysRevA.91.033628,barankov2006synchronization}, where the decay of amplitude of Higgs oscillations was inspected while the uniformity of the system was maintained. 

The relation between frequency of the small amplitude Higgs mode and strength of the pairing correlations ($\hbar\omega = 2\Delta_0$) makes it a very valuable tool from the perspective of experimental measurements. For this reason, recent works (such as~\cite{arxiv.2210.09829}) focus on the stabilization of this mode. Here, and also in~\cite{Dzero2009CooperPT}, we demonstrate that the nonuniform state emerging from the decay of the mode can provide access to new class of states with order parameter that is periodically modulated in space. 
Such states are frequently associated with exotic types of superfluidity, generically related to Fulde-Ferrell-Larkin-Ovchinnikov phase -- a state that is routinely discusses in the context of spin-imbalanced systems. Indeed, when introducing a spin-imbalance to the configuration, the states emerging from the decay consist of many spin-polarized droplets~\cite{PhysRevA.100.033613,arxiv.2211.01055}. Surprisingly, experiments devoted to Higgs modes in strongly interacting Fermi gases may contribute to deeper understanding of exotic states emerging in superfluids.  

\respadd{Tunable ultracold atomic gases confined in box traps can be potentially used to verify the findings of this work~\cite{Navon2021}. Such setups have already demonstrated their high capabilities for detecting the system's nonuniformities. For example, there were successfully used to visualize density modulations due to the propagation of sound waves up to wavelengths of the order 0.1 $\kF$~\cite{Patel2020} -- the resolution that should be sufficient to detect predicted here inhomogeneities in the post-decayed state. }

\vspace{1cm}
\paragraph*{Author contributions}
ABa, ABo and GW contributed to the calculations and data analysis. 
All authors contributed to research planning, interpretation
of the results and manuscript writing.

\paragraph*{Funding information}
This work was supported by the Polish National Science Center (NCN) under Contracts No. UMO-2017/26/E/ST3/00428 (GW, ABa, ABo) and UMO-2021/43/B/ST2/01191 (PM). 
The calculations were executed on Piz Daint supercomputer based in Switzerland at Swiss National Supercomputing Centre (CSCS), PRACE allocation No. 2021240031. We also acknowledge the Global Scientific Information and Computing Center, Tokyo Institute of Technology for resources at TSUBAME3.0 (project ID: hp220072).

\paragraph*{Data availability}
The datasets used and/or analysed during the current study available from the corresponding author on reasonable request.
In order to reproduce numerical results presented in this paper, use reproducibility packs of W-SLDA Toolkit, provided as the supplementary material~\cite{SM}. They contain detailed information about the numerical setup and computation process.

\bibliographystyle{ieeetr}

\end{document}